\documentclass[aps,prl,twocolumn,superscriptaddress,showpacs]{revtex4}

\usepackage{graphicx}
\usepackage{mathrsfs}
\usepackage{bm}
\usepackage{verbatim}
\begin{document}
\title{Transport discovery of emerging robust helical surface states in  $Z_2=0$ systems}
\author{Hua Jiang}
\affiliation{Department of Physics and Jiangsu Key Laboratory of Thin Films, Soochow University, Suzhou 215006, China }
\author{Haiwen Liu}
\affiliation{International Center for Quantum Materials, Peking University, Beijing 100871, China}
\affiliation{Collaborative Innovation Center of Quantum Matter, Beijing, China}
\author{Ji Feng} \email{jfeng11@pku.edu.cn}
\affiliation{International Center for Quantum Materials, Peking University, Beijing 100871, China}
\affiliation{Collaborative Innovation Center of Quantum Matter, Beijing, China}
\author{Qingfeng Sun}
\affiliation{International Center for Quantum Materials, Peking University, Beijing 100871, China}
\affiliation{Collaborative Innovation Center of Quantum Matter, Beijing, China}
\author{X.C. Xie}\email{xcxie@pku.edu.cn}
\affiliation{International Center for Quantum Materials, Peking University, Beijing 100871, China}
\affiliation{Collaborative Innovation Center of Quantum Matter, Beijing, China}

\date{\today}
\begin{abstract}
We study the possibility of realizing robust helical surface states in $Z_2=0$ systems.  We find that the combination of anisotropy and finite-size confinement leads to the emergence of robust helical edge states in both 2D and 3D $Z_2=0$ systems.  By investigating an anisotropic Bernevig-Hughes-Zhang model in a finite sample, we  demonstrate that the transport manifestation of the surface states is robust against non-magnetic disorder, resembling that of a $Z_2 = 1$ phase. Notably, the effective energy gap for the robust helical  states can be efficiently engineered, allowing for potential applications as valley filters and  valley valves.  The realization of emerging robust helical surface states in realistic material is also discussed.

\end{abstract}
\date{\today}
\maketitle
\textit{Introduction.---} 2D quantum spin Hall effect (QSHE) and 3D strong topological insulators (STIs), characterized by the time reversal (TR) invariant $Z_2=1$, have  generated extensive interests in recent years~\cite{CLKane1,CLKane2}.  The hallmark of these novel phases is the existence of odd pairs of helical edge (2D) or surface (3D) states that robust against TR conserving perturbations. %The unusual properties of surface states directly result exotic  phenomena.
In 2D, the robust helical edge states give rise to quantized local and nonlocal conductance~\cite{SCZhang2,SCZhang3,RRDu} and spin polarized edge current~\cite{Molenkamp}. In 3D, the robust helical surface states with spin-momentum locked gapless dispersion~\cite{HJZhang,Hasan} lead to half-integer quantum Hall effect~\cite{Sasagawa}, weak anti-localization~\cite{YQLi}, absence of backscattering~\cite{RJCava}  etc~\cite{CLKane1}.  These exotic properties make their host systems an ideal  platform for testing  fundamental physical paradigms and promising application in low-power dissipation information processing.

However, the requirement of $Z_2=1$ for the existence of robust helical surface/edge states is rather stringent. Indeed, the QSHE is only experimentally confirmed in  HgTe/CdTe and InAs/GaSb quantum wells~\cite{SCZhang2,RRDu}. The scarcity of host systems represents a materials challenge, hindering the study and development  of devices based on robust helical edge states. In 3D systems, one may note other classes of TIs that harbor multiple Dirac surface states:  $Z_2=0$ weak topological insulators (WTIs) ~\cite{LiangFu,BHYan,PZTang,WHDuan,LiXiao} and topological crystalline insulators ~\cite{LiangFu1,Hsieh,Buczko,XuSY,Tanaka,Dziawa}.  More recently, a WTI material ${\rm Bi_{ 14}Rh_{ 3}I_{ 9}}$ is  successfully fabricated in experiment and generate intensive attraction~\cite{BRasche}.
%In general, s WTI material has even pairs of helical surface states, and their topological properties are  studied~\cite{LiangFu,KIImura1,RSKMong,ZRingel,KIImura2,KIImura3}.
It becomes highly desirable to be able to engineer these materials with multiple Dirac states into structures with robust topological transport.

In this Letter, we show the emergence of robust helical edge (surface) states in  both 2D and 3D $Z_2=0$ systems, arising from anisotropic confinement in a finite-size sample. Based on transport simulations of anisotropic  Bernevig-Hughes-Zhang (BHZ) model, we demonstrate quantized conductance of helical edge states under strong nonmagnetic disorders. The robustness of helical surface states due to anisotropic confinement is generalizable to 3D WTIs. Moreover, the proposed $Z_2=0$ systems possess  additional exotic properties than in $Z_2=1$ TIs.  In particular,  by controlling the sample size and strain engineered anisotropy, this mechanism allows for  efficient tuning of the effective energy gap, and  fabrication of valley filter and valley valve without breaking TR symmetry.  Finally,  the two realistic material systems that host emerging robust helical surface states are proposed.

\textit{2D model.---}  We consider an anisotropic BHZ model in a square lattice ~\cite{Bernevig,XLQi}. The four-band tight-binding Hamiltonian in the momentum representation reads:
\begin{eqnarray}
\mathcal{H}(\vec{k})&=& \left(
                          \begin{array}{cc}
                            h(\vec{k}) & 0 \\
                            0 & h^{*}(-\vec{k}) \\
                          \end{array}
                        \right),  \nonumber\\
h(\vec{k}) & = & \tau_z (m-m_x -m_y + m_x  \cos k_x+ m_y  \cos k_y)  \nonumber\\
&+& \tau_x \upsilon_x  \sin k_x + \tau_y \upsilon_y  \sin k_y
,\label{Equation1}
\end{eqnarray}

where $h(\vec{k})$ and its time reversal  $h^{*}(-\vec{k})$ are, respectively, decoupled Hamiltonians for the two spins.
The Pauli matrices, $\vec{\tau}$, address the orbital space. This model involves five parameter. Specifically, m determines the band gap, $v_{x,y}$ reflects the Fermi velocity, and $m_{x,y}$ represent the hopping amplitudes between nearest-neighbor sites along $x/y$ direction, respectively. The gap parameter, m, is a key variable in subsequent simulations. We adopt the following values for the other parameters unless otherwise specified: $m_x=0.8,m_y=1.2,\upsilon_x=\upsilon_y=3.0$.

\begin{figure}
\includegraphics[width=8cm,  viewport=12 622 575 832, clip]{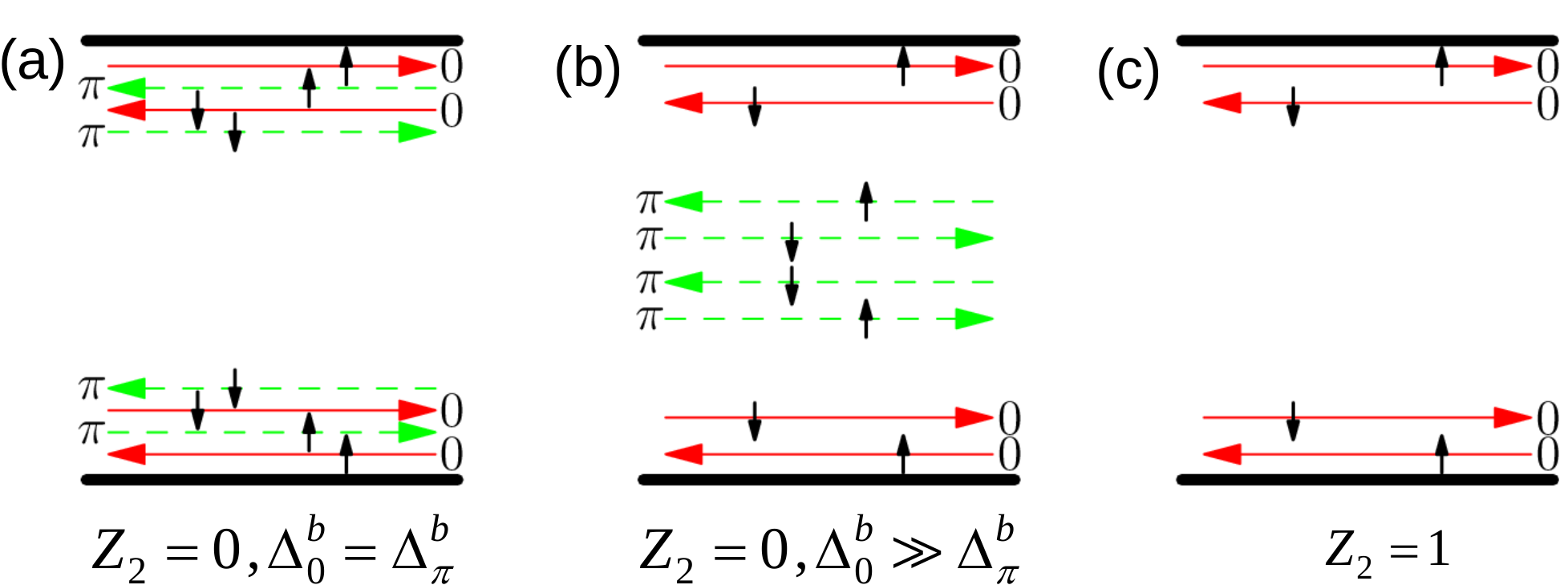}
\caption{(Color online)
(a)-(c) Schematic plot of helical edge modes  in the anisotropic BHZ stripe with y-termination under  parameter $m = 2.00, ~1.64, ~1.56$. The vertical arrows represent electron spin.  In subplot (b), the helical edge channels around $k_x=\pi$ are hybridized due to finite size confinement, leaving one pairs of  helical edge channels around $k_x=0$, thus these conducting channels are resemble to subplot (c).
 }\label{Fig1}
\end{figure}

\begin{figure*}
\includegraphics [width=15.5cm, viewport=8 233 820 538, clip]{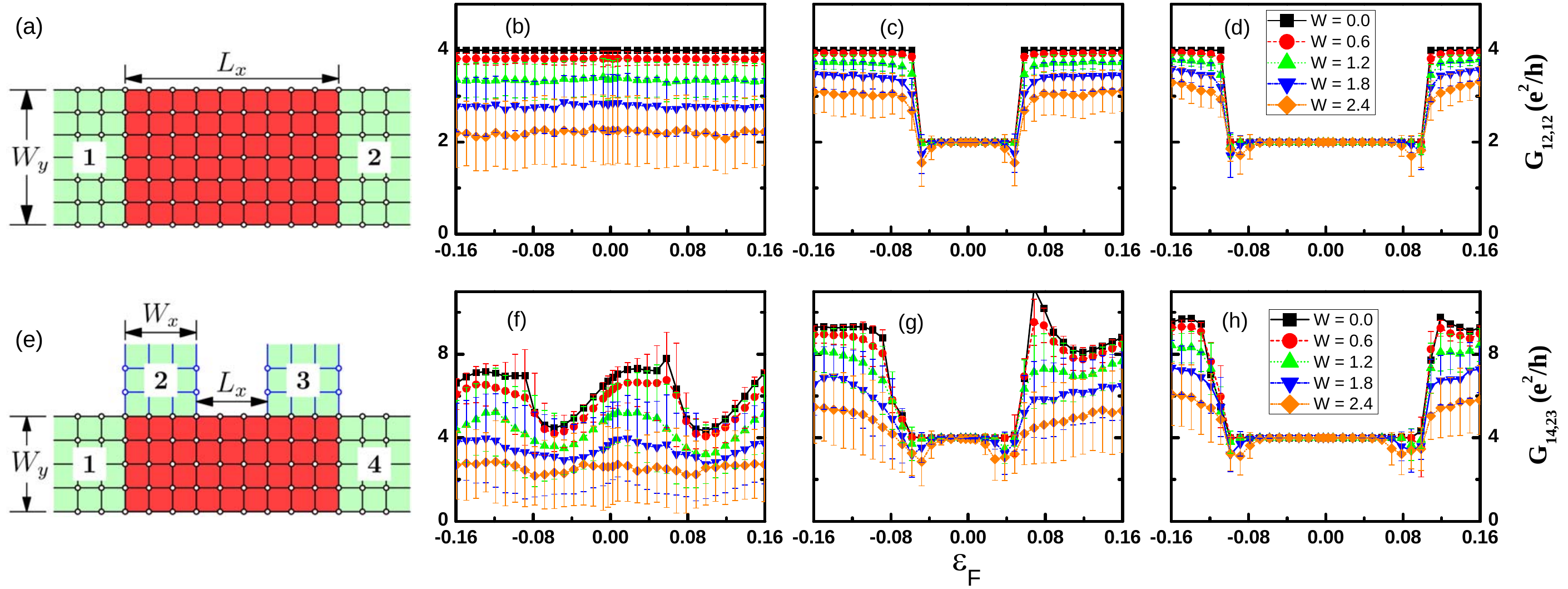}
\caption{(Color online) (a)(e) Schematic plot of two-terminal  and  $\pi$ bar devices . The disorder are  considered in the central (red filled) region. The size parameters are $L_x =120$, $W_x=120$, $W_y=60$. (b)-(d)  are the two terminal conductance $G_{12,12}$ of device (a), and  (f)-(h) are nonlocal conductance $G_{14,23}$ of device (e) versus Fermi energy ${\rm \varepsilon_F}$ for different disorder strength $W$ at various $m=2.00$ (b)(f), $~1.64$ (c)(g), $~1.56$ (d)(h) . The error bars denote the conductance fluctuation. }\label{Fig2}
\end{figure*}

We first examine the  topological properties of bulk phase described by this model, by calculating the $Z_2$ invariant~\cite{LiangFu}. A $Z_2 =1$ QSHE is obtained when $m \in (0,2m_x) \cup (2m_y, 2m_x +2 m_y)$ ~\cite{Bernevig,XLQi}. In contrast, $Z_2=0$ when  $m \in (-\infty, 0)  \cup (2m_x + 2m_y, \infty)$ or $m \in (2m_x, 2m_y)$. However, the two regions with $Z_2=0$ are distinct from each other. When $m  \in (-\infty, 0)  \cup (2m_x + 2m_y, \infty) $, all bands are normal bands as ordinary insulator, whereas region $m \in (2m_x, 2m_y)$ is non-trivial in the case of anisotropic $m_x \neq m_y$. It contains two inverted bands around $k_x=0$ and $\pi$, see supplementary  for details~\cite{supple}. The inverted bands  guarantee the existence of helical edge modes, whose amplitude decays from y-boundary to the bulk exponentially with a decay length  $\xi \sim 2\upsilon_y/\Delta^b$. Here, $\Delta^b$ denotes the inverted bulk gap~\cite{HZLu}.  When $m \in (2m_x, 2m_y)$, the two inverted gaps ($\Delta_0^b=4m_y-2m$ at $k_x=0$, $\Delta_\pi^b=2m-4m_x$  at $k_x =\pi$) result in two pairs of helical edge modes labeled by $0$ and $\pi$, respectively.

Narrow width or thickness of a $Z_2=1$ TI is often viewed to impose adverse effects on the transport of helical surface states. In a 2D system,  the edge channels on the two sides of a narrow sample becomes hybridized, leading to undesirable backscattering  between  edge states  by weak disorder~\cite{HZLu}.  In a 3D system, the high-quality samples are always grown via layer-by-layer MBE techniques. It was indeed observed in ultra-thin films that the finite size effect leads to hybridization gap, destroying the robustness of helical surface state and drastically diminishing the surface conductivity~\cite{QKXue}.

Key to our proposal is the observation that in a finite sample, these two helical edge states behave differently for various $m$ [see Figs. 1(a) - (b)]. Intriguingly, when $m$ approaches the critical value  corresponding to the topological phase
transition (i. e. $m=1.64$),   the bulk gap at $\pi$ approaches zero, $\Delta^b_\pi\rightarrow 0$, while $\Delta^b_0$ remains finite. Consequently, the decay length
$\xi_\pi$ of $\pi$ helical edge channels is exponentially long, while $\xi_0$ of  $0$ helical edge channels can in principle be much smaller than sample width. The strong hybridization of $\pi$ helical edge states annihilates the corresponding edge channels, whereas the small $\xi_0$ guarantees the survival of $0$ helical edge channels. Therefore, from a transport point of view as shown in fig. 1(b), the conducting edge channels are similar to Fig. 1(c), although their $Z_2$ invariants defined for the bulk systems are different. Besides, there are two pairs of conducting channels
in Fig. 1(a), despite the fact that $Z_2=0$.  The similarity between conducting edge channels between the $Z_2=0$ system [fig. 1(b)] and the $Z_2=1$ system [fig. 1(c)] leads naturally to the speculation that the emergent $k_x=0$ helical edge states are as robust as the edge states in  a $Z_2=1$ system .

In order to quantitatively assess the robustness of those emerging $Z_2=0$ helical edge states, we inspect a two-terminal device and a  $\pi$-bar device [see Figs. 2(a),2(e)],  and  study their transport properties in the presence of TR-conserving disorder using the  Landauer-B\"{u}ttiker formula~\cite{SDatta,HJiang1,HJiang2}. The longitudinal terminals are perfect leads with the same parameters as the  central region, and the transverse terminals are metallic leads. The disorder is modeled by the Anderson-type random  on-site potential  uniformly distributed in the range $[-\frac{W}{2},\frac{W}{2}]$.  As in the experimental setup~\cite{SCZhang2,RRDu},  two terminal conductance $G_{12,12}$ and nonlocal conductance $G_{14,23}$ of these two  devices are systematically assayed in our simulations.

In Figs. 2(b)-2(d), two-terminal conductance $G_{12,12}$ and the corresponding fluctuation versus Fermi energy $\varepsilon_F$ under various  $m$ are plotted.
The case of emergent helical states with $m=1.64$ is shown in Fig. 2(c).
The conductance $G_{12,12}$  shows two quantized plateaus $\frac{4e^2}{h}$ and $\frac{2e^2}{h}$ in the clean limit. In the presence of strong disorder, the $\frac{4e^2}{h}$  conductance decreases rapidly while $\frac{2e^2}{h}$  plateau remains unchanged with vanishing fluctuation.  Therefore, the two-terminal transport properties of emergent helical edge states in $Z_2=0$ case  behave exactly like a $Z_2 =1$ QSHE, as shown in Figs. 2(c)-(d). The conductance is completely different from the  $Z_2=0,\Delta^b_0=\Delta^b_{\pi}$ case,  where only $G_{12,12}=\frac{4e^2}{h}$ plateau is present in the clean limit, which is fragile against disorder is observed [see fig.2(b)]. This is because the carriers can scatter between the counter-propagating 0 and $\pi$ channels in this case [see fig. 1(a)]. In contrast, due to the vanishing  of $\pi$ helical edge states under finite size confinement, the emergent helical states described above are not susceptible to these scattering channels.

The two-terminal measurements can also be corroborated by $\pi$-bar measurements [see Figs. 2(f)-2(h)]. For the case with emergent helical edge states,  $G_{14,23}$ demonstrate well quantized plateaus at $\frac{4e^2}{h}$, irrespective of leads detail and strong disorder strength [see fig. 2(g)]. The two-terminal perfect $\frac{2e^2}{h}$ plateau and $\pi$-bar perfect $\frac{4e^2}{h}$ conductance plateaus should plausibly lead to a transport definition of robust helical edge states in these $Z_2 = 0$ systems. In a $Z_2=1$ topological insulator, the robustness of the edge conduction is derived from its intrinsic topological character. The robust helical edge states in our $Z_2=0$ model, on the other hand, the intervalley scattering is detuned by the hybridization gap of one of the valleys.

Besides the emergence of robust helical edge states in $Z_2=0$ system, which conveniently augments the scope of study confined in QSHE, our proposal also has
certain adavantages over the helical states in $Z_2=1$ systems.
For example, the energy window with robust edge channel  in our proposed model can be easily engineered via tailoring the sample width, while such window in $Z_2=1$ system is difficult to change in experiments. Figure 3 plots two terminal conductance $G_{12,12}$ as a function of ${\rm \varepsilon_F}$ under various sample width $W_y$ for these two systems.  With increasing $W_y$, in $Z_2=0$ system the  energy window with robust $\frac{2e^2}{h}$  plateau continuously decreases from a moderate value to zero  [see Figs. 3(a) and 3(b)]. In contrast, the  window of plateau in $Z_2=1$ system is basically equal to bulk  gap and insensitive to the width variation [see fig. 3(c) and 3(d)].  To be specific, the energy window, which is appropriately termed an effective energy gap, in Fig. 3(a) arises from the hybridization  gap of the $\pi$ helical edge states, and can be continuously tuned by tailoring the sample width [see fig. 3(b)].

\textit{3D model.---} The similar  phenomena also exist in 3D anisotropic WTI.
We take the anisotropic Wilson-Dirac-type model as an example~\cite{CXLiu,KIImura1,RSKMong,ZRingel,KIImura2,KIImura3}. The Hamiltonian in cubic lattice reads:
\begin{eqnarray}
\mathcal{H}(\vec{k}) &=& m (\vec{k}) \sigma_z \otimes \tau_0 +
\sum_\alpha \upsilon_\alpha \sin k_\alpha ~\sigma_\alpha \otimes \tau_x ,\nonumber\\
m (\vec{k})&=& m + \sum_\alpha m_\alpha  (\cos k_\alpha-1),
\label{Equation2}
\end{eqnarray}
%where $\sigma$ and $\tau$ are Pauli matrices in spin and orbital spaces. $m$ represent gap parameter. $\upsilon_\alpha, m_\alpha$ denote Fermi %velocity and  hopping energy along $\alpha=x,y,z$ direction.
where $\sigma$ are the Pauli matrices in spin spaces, and $\alpha = x,y,z$. Parameters $\tau, m, \upsilon_\alpha, m_\alpha$ have the same meanings as in the 2D model of Eq. \ref{Equation1}.
Recently, several works have studied the  finite size effect and transport properties of 3D WTI~\cite{KIImura1,RSKMong,ZRingel,KIImura2,KIImura3}. These works focus on two cases: (i) isotropic bulk, i.e. $m_x=m_y=m_z$; (ii) anisotropic bulk but isotropic xy surface, i.e. $m_x=m_y \neq m_z$.  Interestingly, we find the combined effects of finite size confinement and  anisotropic surface, i.e. $m_x \neq m_y$, leads to unique phenomenon in $xy$ surface.
The anisotropy may be induced by elastic strain engineering. For example, when an isotropic WTI film is deposited to a substrate with uniform tensile (compressive) strain along $x$ direction, the anisotropic WTI with $m_x < m_y$  ($m_x > m_y$) is obtained [see, respectively, regions II and IV in Fig. 4(d)].
Henceforth, both tensile and compressive strains are explored in terms of relative values of $m_x$ and $m_y$, while other parameters are fixed: $m=2.26,m_z=0.8$, $\upsilon_x=\upsilon_y=\upsilon_z=1.5$ .
%In all cases, the system belong to the WTI with $Z_2$ indices $(0,111)$.

\begin{figure}
\includegraphics[width=8cm,  viewport=25 400 575 762, clip]{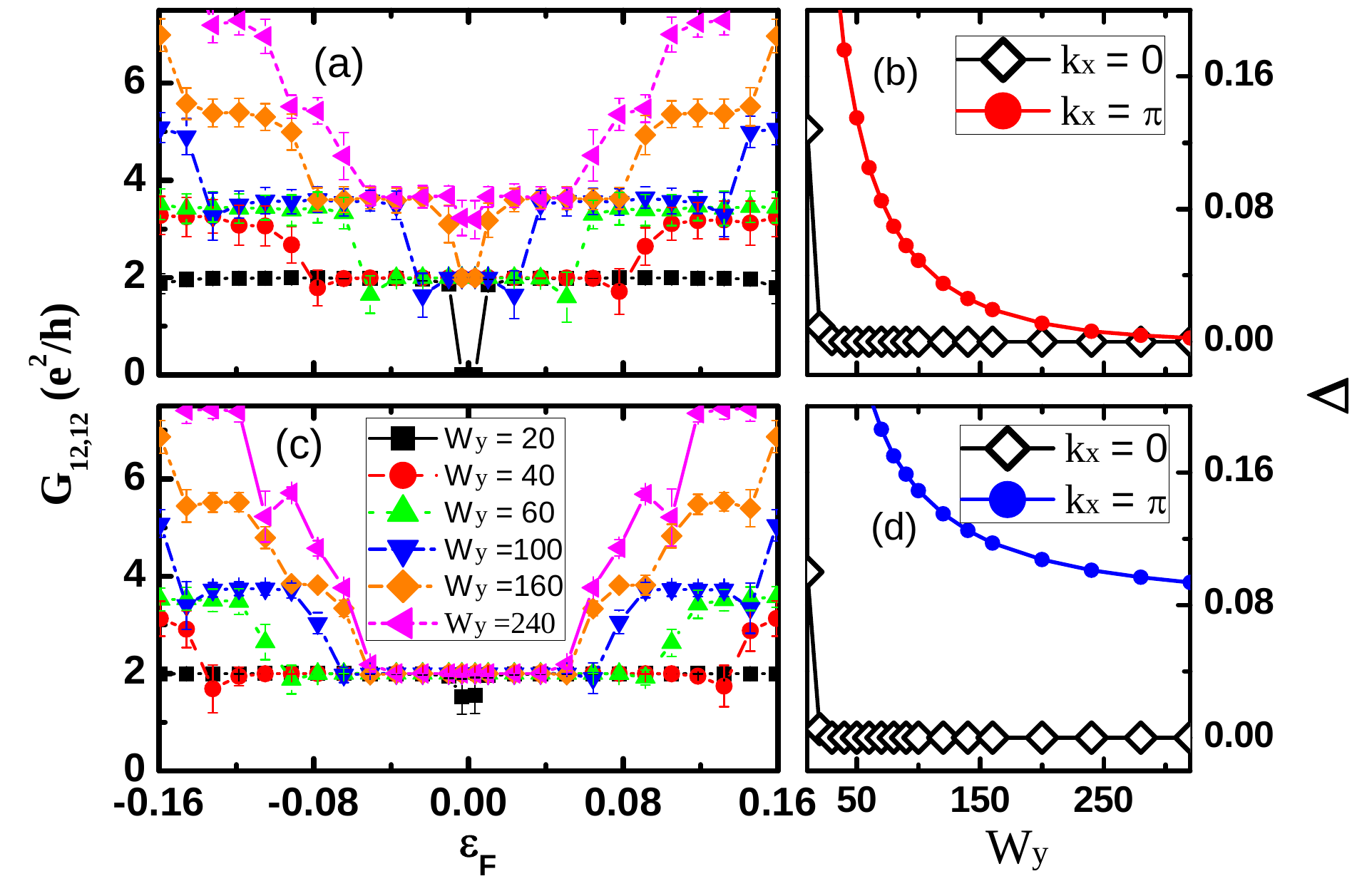}
\caption{(Color online) (a)(c) The two terminal conductance  $G_{12,12}$ versus Fermi energy $\varepsilon_F$ for  different sample width $W_y$ at various $m=1.64$ (a),  $1.56$ (c), with   $L_x =120$ and  $W=1.8$.  (b)(d) show finite size energy gap $\Delta$ of $k_x = 0 (\pi)$ as a function of $W_y$ at fixed $m=1.64$ (b) and $1.56$ (d).      }\label{Fig3}
\end{figure}

\begin{figure}
\includegraphics [width=8.5cm, viewport=8 257 610 818, clip]{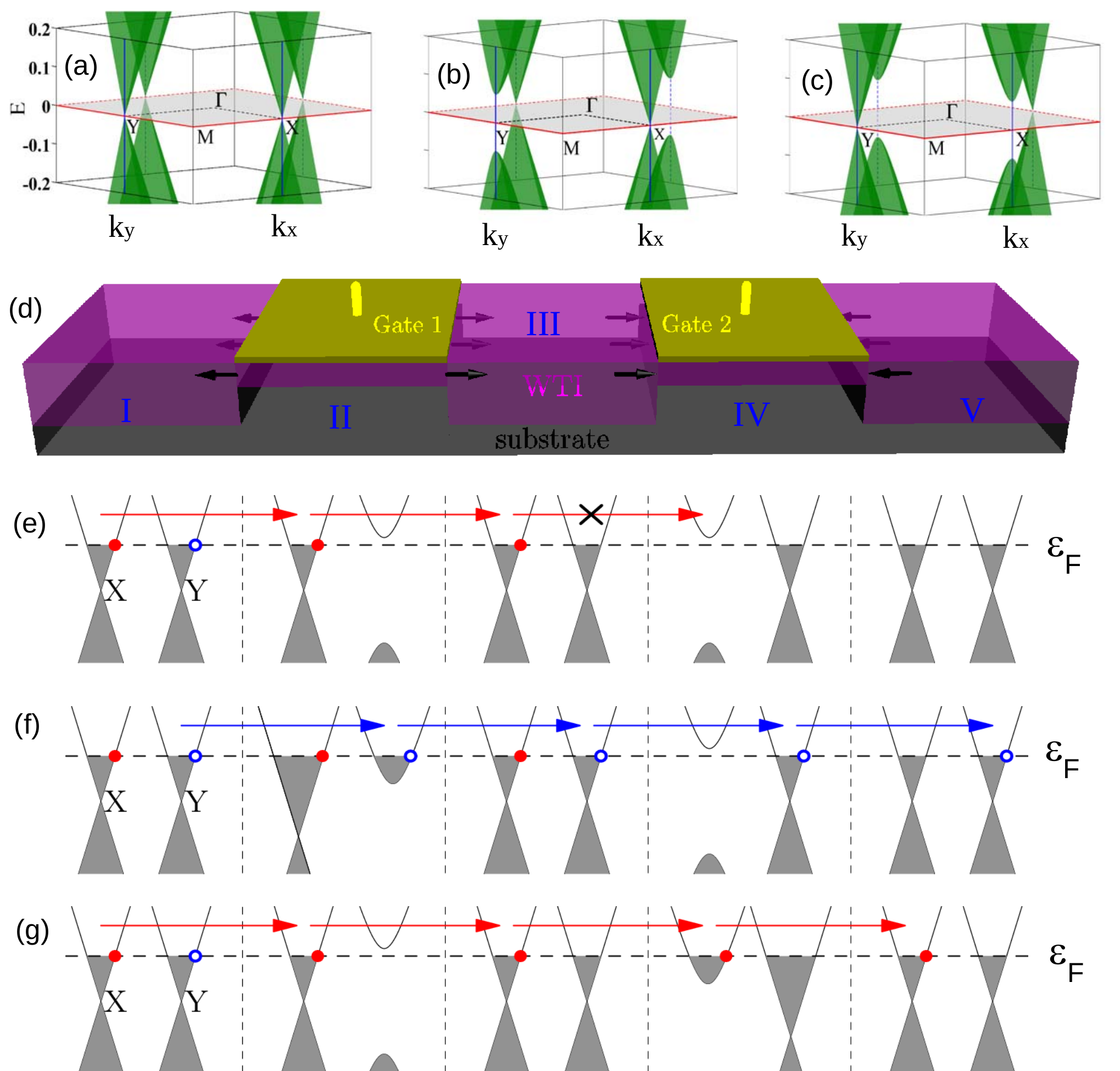}
\caption{(Color online) (a)-(c) The surface energy bands  of anisotropic WTI film with parameters $m_x=0.9, m_y =1.1, {\rm N_z} =120$ (a), $m_x=0.9, m_y =1.1, {\rm N_z} =20$ (b), $m_x=1.1, m_y =0.9, {\rm N_z} =20$ (c).  ${\rm N_z}$ is the sample thickness.  %The system belong to the WTI with $Z_2$ indices $(0,111)$.
(d) Schematic diagram of the valley filter and valley valve device. The Fermi energy in region II (IV) can be tuned by the attached gates.  The black arrows with in (out) direction represent the tensile (compress) strain. (e)-(h) Schematic plot of the working mechanisms for the device. Open (filled) circle  denotes the electrons in X (Y) valley. }\label{Fig4}
\end{figure}

Figs. 4(a)-(c) schematically display the evolution of  surface energy bands of  anisotropic WTI film. For sufficiently thick films,
two gapless helical surface modes with Dirac cone ${\rm X}(0, \pi)$ and ${\rm Y}(\pi,0)$  exist at the top surface [see fig. 4(a)]. ${\rm X}$, ${\rm Y}$ are two valleys analogue to K and K' in Graphene.
In thinner films, the finite size confinement leads to hybridization gaps in helical surface states.  As shown in Fig. 4(b), the valley ${\rm X}$ remains almost gapless while hybridization gap  $\Delta^{s}_{\rm Y}$ at valley ${\rm Y}$ is remarkable. This peculiar feature arises from coupling strength of  surface state at $\rm X$ is much weaker than that at $\rm Y$, which is due to bulk inverted gaps satisfying $\Delta_{\rm X}^b=2m-4m_x \gg \Delta_{\rm Y}^b=2m-4m_y$ in tensile strained WTI film. Similar to the 2D case, when $\varepsilon_F$ is located inside the hybridization gap of ${\rm Y}$ valley, the ${\rm X}$ helical surface states survive, and intervalley scattering is definitely avoided. In other words, the emerging ${\rm X}$ helical surface states are robust and share the common features of those states in STIs. Moreover, the energy window (effective energy gap) with robust conducting states can  be engineered via varying the sample thickness. Specifically, the gapped ${\rm Y}$ valley case can be tuned into the gapped  ${\rm X}$ valley case by changing the system from tensile strain to compress strain [see Fig. 3(c)].

The degree of valley polarization provides promising route towards potential quantum applications ~\cite{CWJBeenakker,DiXiao,JFeng,JIsberg}.
In typical valleytronics materials, e.g. graphene, MoS$_2$ etc, the band gaps of distinct valley are related by discrete symmetry (such as time-reversal), making it difficult to create valley valve and filter  devices. In anisotropic WTIs, the two valleys are not interrelated by any discrete symmetry. This unique feature provides experimentally feasible pathway to independent tuning of the gaps at each valley, allowing for construction of interesting valley devices. Below, we propose a valley filter and valley valve device based on anisotropic WTI, as illustrate in fig. 4(d). The region I and V are source and drain, which can be fabricated by either isotropic or anisotropic thick WTI . The region II and IV are fabricated by thin anisotropic WTI films with tensile  and compress strain,  respectively.
%The Fermi energy of region II and IV can be easily tuned by the attached gates due to its film feature.
The region III, made by thick WTI,  can relax strain from region II to IV.  Figs. 4(e)-4(g) illustrate the working mechanism of the device. The valley valve is  illustrated in fig. 4(e). Due to  momentum mismatch,  the electron tunneling  in {\rm Y} valley is forbidden, and the survived X valley electrons in region III can hardly tunnel  to region V.  Further, the only biased  gate 1  leads to a ${\rm Y}$ valley filter [see Fig. 4(f)]. Both  ${\rm X}$ and ${\rm Y}$ valley electrons can tunnel from region I to III while ${\rm Y}$ valley electrons are elected after region IV.  Similarly,  a ${\rm X}$ valley filter is obtained by  only biased gate 2, as shown in fig. 4(g).  Moreover,  by appropriately tuning the bias of two saperate gates, a $100 \%$ {\rm X} valley polarization can be continuously switched to $100 \%$ {\rm Y} valley polarization. To emphasize, this proposed device has two advantages: (i) the
complete valley manipulation can be obtained by current experimental electric techniques;  (ii) the magnitude of valley current is remarkable with low dissipation,  protected by the robust transport properties of helical surface states. 

\textit{Material discussion ---} The proposed model with emerging robust helical surface states could in principle be realized and probed in  realistic $Z_2 = 0$ materials. Two candidate materials are noteworthy ~\cite{supp1}. (1) Bismuth (111) film with a thickness between 20 and 70  nm, where the robust helical surface states lead to several experimental observable phenomena, namely, the fact that surface bands can cross Fermi level an odd number of times between two time-reversal invariant points, and consequently, weak anti-localization behavior. Moreover, this prediction is also highly relevant to an unpublished experiment~\cite{XFJin}. (2) The ${\rm (SnTe)_7(CaTe)_1(110)}$ superlattice film with $0.1\%-0.4\%$ uniform c-axis tensile strain, with the thickness range of about 100~{\rm nm} to 200~{\rm nm} for a given strain ratio~\cite{supp1}.

\textit{Summary.---} We demonstrate the emergence of robust helical edge (surface) states in both  2D  and 3D anisotropic $Z_2=0$ systems. The anisotropy and finite size confinement play important roles in realizing such states. These emerging  robust helical states lead to the revival of major features of  $Z_2=1$ systems.  In addition, the effective energy gap for  robust helical states can be  engineered by tailoring sample size.  These characteristics have potential application for valley filter and valley valve under current experimental techniques.

\textit{Acknowledgements.---} We are grateful to Q. Niu, S. Q Shen,  L. Fu and especially X. F. Jin for valuable discussions.  This work was financially supported by NBRPC (2009CB929100, 2011CBA00109, 2012CB921303, 2012CB821402 and 2013CB921900),  NSFC (91221302, 11274364, 11174009 and 11374219) and CPSF(2013T60020).

\end{document}